\documentstyle[11pt]{article}

\title{\bf Double scattering on the nucleus in the perturbative QCD}
\author{M.Braun \\ Department of High Energy physics,
 University of S.Petersburg,\\
198904 S.Petersburg, Russia}
\def\beq{\begin{equation}}
\def\eeq{\end{equation}}
\def\noi{\noindent}
\begin{document}
\maketitle
\medskip
\noi{\bf Abstract.}

In the hard pomeron model consequences are studied
 which follow from the recently obtained form of the diffractive amplitude
for the double
scattering on the nucleus
and the related EMC effect at small $x$. It is shown that at large $Q^2$
the double scattering contribution to the latter falls as $Q^{-0.6338}$ and
in all probability dominates the total effect.\vspace{1 cm}

{\Large\bf SPbU-IP-1998-1}

\newpage

\section{Introduction. Double scattering}
Perturbative QCD predicts that hadronic scattering at high energies
proceeds via the exchange of a hard (BFKL) pomeron, with an intercept
above unity. Much effort has been spent in analysing possibilities
to see a signature of the hard pomeron in the observed behaviour of the
structure functions at low $x$  and in jet production.

Recently a new object has drawn  considerable attention within the
perturbative approach, namely, the diffractive amplitude and the triple
pomeron interaction related to it [1-4]. The found diffractive amplitude
turned out to have a rather specific form. Since this amplitude
governs the double scattering contribution to the cross-section on the
nucleus, these findings open new possibilities to compare predictions
of the perturbative approach with the experiment.

In short, the diffractive amplitude calculated perturbatively 
 and in the high-colour limit has been found to possess the following
 properties [1,4]

 1. The double pomeron exchange contribution to it can be completely
 absorbed into the triple pomeron one.

 2. At small momentum transfers $\kappa$ the latter is singular.

 3. At small $\kappa$ the coupling of the pomeron to a hadronic
 projectile behaves as $\ln\kappa$

These properties have been established in a pure perturbative approach,
with a fixed strong coupling constant $\alpha_s$. It is not known in
what manner they will change if one introduces confinement effects in
any form.

In this note, rather than speculate on the latter problem, we study
 the observable consequences of these properties for the double scattering
 on the nucleus. To stay within the perturbative appproach the
 projectile has to be either a heavy "onium" or a highly virtual photon.
 Having in mind applications to the low $x$ EMC effect, we choose the
 latter choice.

 The amplitude ${\cal A}_2$ corresponding to the double scattering
 on the nucleus and normalized according to $2\,{\mbox Im}{\cal A}
 =\sigma^{tot}$
 is given by the well-known expression (for $A>2$)
 \beq
{\cal A}_2=iC_A^2\int\frac{d^2\kappa}{(2\pi)^2}F_A(\kappa)a_2(\kappa)
 \eeq
 Here $F_A(\kappa)$ is a two-nucleon form-factor of the nucleus:
 \beq
 F_A(\kappa)=\int d^2r_1d^2r_2\rho_A(r_1,r_2)e^{i\kappa(r_1-r_2)},\ \
 \kappa_z=0
 \eeq
 where $\rho_A$ is a two-nucleon density. If it is taken factorized
 (no correlation approximation) then
 \beq
 F_A(\kappa)=T_A^2(\kappa)
 \eeq
 where $T_A(\kappa)$ is a two-dimensional Fourier transform of the
 standard nuclear profile function $T_A(b)$. For the deuteron, instead of (2),
 \beq
 F_2(\kappa)=\int d^2r|\psi(r)|^2e^{i\kappa r}, \ \
 \kappa_z=0
 \eeq
 where $\psi$ is the deuteron wave function.

 The high-energy part $a_2(\kappa)$ is given by an integral
 \beq
 ia_2(\kappa)=\frac{1}{8\pi s^2}\int_0^{\infty}ds_1
 {\rm Disc}_{s_1}D(s,s_1,\kappa)
 \eeq
 where $D(s,s_1,\kappa)$ is a diffractive amplitude for the c.m. energy
 squared $s$, diffractive mass squared $s_1$ and transferred momentum
  $\kappa$.

  We recall that the eikonal (Glauber) approximation consists in taking in (5) only
  a contribution from the intermediate state equal to the initial
  projectile hadron, which gives $a_2$ independent of $\kappa$:
  \beq
  a_2(\kappa)=a^2
  \eeq
  where $a$ is the forward projectile-nucleon amplitude
  (also normalized to $2{\rm Im}\, a=\sigma^{tot})$. Then (1) and (3)
  immediately lead to the well-known result
  \beq
 {\cal A}_2=iC_A^2 a^2 w_2,\ \ w_2=\int d^2b T_A^2(b)
  \eeq

 \section{Triple pomeron contribution}
  The diffractive amplitude found in the perurbative QCD is
  totally given by a contribution from the triple pomeron interaction.
  So the corresponding double scattering amplitude on the nucleus
  will have an essentially non-Glauber form.

  The discontinuity entering (5) turns out to be given by
\[
D_1\equiv\frac{1}{2i}{\rm Disc}_{s_1}D(s,s_1,\kappa)
=\frac{\alpha_s^5}{\pi^3}N^{2}(N^{2}-1)\frac{s^{2}}{s_{1}}
\int \prod_{i=1}^{3}d^{2}r_{i}\frac{r_1^2\nabla_1^4}{r_2^2r_3^2}\]
\beq\exp (-i\kappa(r_{2}+r_{3})/2)
\phi_{1}(s_{1},0,r_{1})\phi_{2}(s_{2},\kappa,r_{2})
\phi_{2}(s_{2},-\kappa,-r_{3})\delta^2(r_1+r_2+r_3)
\eeq
Here $\phi_i(s,\kappa,r)$, describe the upper ($i=1$) and two lower
pomerons ($i=2$, see Figure) for the energetic variable $s$, momentum
$\kappa$ and the intergluon transverse distance  $r$. $N$ is the
number of colours; $s_2=s/s_1$. The pomerons are
joined by the triple pomeron vertex, whose form was obtained in [4] in the
high-colour limit. Presenting the $\delta$-function as an integral over
the auxiliary momentum $q$ we rewrite (8) in the form
\beq
D_1=\frac{\alpha_s^5}{\pi^3}N^{2}(N^{2}-1)\frac{s^{2}}{s_{1}}
\int \frac{d^{2}q}{(2\pi)^2}\chi_{1}(s_{1},0,q+\kappa/2)
\chi_{2}^2(s_{2},\kappa,q)
\eeq
where
\beq
\chi_{1}(s,0,q)=\int d^{2}rr^2\nabla^{4}\phi_{1}(s,0,r)\exp iqr
\eeq
and
\beq
\chi_{2}(s,\kappa,q)=\int d^{2}r r^{-2}\phi_{2}(s,\kappa,r)\exp iqr
\eeq

The solutions $\phi_{1(2)}$ can be obtained by using the Green function
of the BFKL equation for a given total momentum $G_{l}(s,r,r')$. For the
projectile (see [5]):
\beq
\phi_{1}(s,\kappa,r)=\int d^{2}r'(G_{s}(\kappa,r,0)-
G_{s}(\kappa,r,r')\
\rho_{1}(r')
\eeq
Here $\rho_{1}(r)$ is the colour density of the projectile as a function
of the intergluon distance with the colour factor $(1/2)\delta_{ab}$
and $g^{2}$ separated. As mentioned, for the projectile we choose
 a highly  virtual
 photon with $p_{1}^{2}=-Q^{2}\leq 0$, which splits into
$q\bar q$ pairs of different flavours. The explicit form of $\rho$ is
well-known for this case [5]. For the transverse photon
\beq
\rho_{1}^{(T)}(r)= \frac{e^{2}}{4\pi^3}
\sum_{f=1}^{N_{f}}Z_{f}^{2}\int_{0}^{1}d\alpha
(m_{f}^{2}{\rm K}_{0}^{2}(\epsilon_{f} r)
+(\alpha^{2}+(1-\alpha)^{2})\epsilon_{f}^{2}{\rm K}_{1}^{2}
(\epsilon_{f} r))
\eeq
where $\epsilon_{f}^{2}=Q^{2}\alpha (1-\alpha)+m_{f}^{2}$ and $m_{f}$
and $Z_{f}$ are the mass and charge of the quark of flavour $f$.
For the longitudinal photon
\beq
\rho_{1}^{(L)}(r)= \frac{e^{2}}{\pi^3}Q^2
\sum_{f=1}^{N_{f}}Z_{f}^{2}\int_{0}^{1}d\alpha
\alpha^2(1-\alpha)^2{\rm K}_{0}^{2}(\epsilon_{f} r)
\eeq
For the hadronic target, we assume an expression similar to (12) with a colour
density $\rho_{2}(r)$ non-perturbative and its explicit
form unknown. For our purpose it is sufficient to know
that the corresponding mass scale is not large.

The two lower pomerons are in their asymptotic regime. So $\chi_2$ can be
found using an asymptotic expression for the pomeron Green function, in the
same way as in [6], to which paper we refer for the details. One obtains
then
\beq
\chi_{2}(s,\kappa,q)=8s^{\Delta}(\pi/\beta\ln s)^{3/2}F_{2}(\kappa)J(\kappa,q)
\eeq
where $\Delta=(\alpha_sN/\pi)4\ln 2$ is the pomeron intercept,
$\beta=(\alpha_sN/\pi)14\zeta(3)$,
\beq
J(\kappa,q)=\int \frac{d^{2}p}{2\pi}\frac{1}
{|\kappa/2+p||\kappa/2-p||q+p|}
\eeq
and $F_2(\kappa)$ desribes the coupling of the lower pomeron to the target
\beq
\int d^{2}Rd^{2}r\frac{\exp(i\kappa R)r\rho_{2}(r)}{|R+r/2||R-r/2|}=
\pi  F_2(\kappa)
\eeq
At small $\kappa$ $F_2$ diverges logarithmically:
\beq
F_2\simeq -4(R/N)\ln\kappa,\ \ \kappa\rightarrow 0
\eeq
where
 \beq R=\frac{N}{2}\int d^2r r\rho_2(r)
 \eeq
 has a meaning of the average target dimension.

 If one also takes an asymptotic expression for the upper pomeron then
 the whole discontinuity $D_1$ aquires a factorized form [6] corresponding
 to the triple pomeron contribution in the old Regge-Gribov theory:
  \beq
  D_1(s,s_1,\kappa)=-\gamma_1\gamma_2^2(\kappa)
  \gamma_{3P}(\kappa)  P(s_1,0)  P^2(s/s_1,\kappa)
  \eeq
  Here $P(s,\kappa)$ is the pomeron propagator
  \beq
  P(s,\kappa)=2\sqrt{\pi}s^{1+\Delta}(\beta\ln s)^{-\epsilon}
  \eeq
  with
  $\epsilon=1/2(3/2)$
  for $\kappa=0(>0)$. 
  The vertices $\gamma_{1(2)}$ describe the pomeron interaction with
  the projectile (target).  Using (13) and (14) one obtains, neglecting the
  quark masses
  \beq
  \gamma_1=\alpha_s\sqrt{N^2-1}(b/Q)
  \exp\left(-\frac{\ln^2Q}{\beta\ln s}  \right)
  \eeq
  where $Q^2$ is the photon virtuality and for a transverse (T) and
  longitudinal (L) photons
  \beq
  b_T=\frac{9\pi e^2Z^2}{256},\ \ b_L=\frac{2}{9}b_T
  \eeq
  $Z^2=\sum_fZ^2_f$.
  The target vertex $\gamma_2$ is proportional to $F_2$, Eq. (17)
  \beq
  \gamma_2(\kappa)=\alpha_s\sqrt{N^2-1}F_2(\kappa)
  \eeq
  and
  so logarithmically  divergent at small $\kappa$.
  The triple pomeron coupling $\gamma_{3P}$ is also
  singular at small $\kappa$:
  \beq
  \gamma_{3P}(\kappa)=\frac{32\alpha_s^2N^2}{\sqrt{N^2-1}}
  \frac{B}{\kappa}
  \eeq
  where the number $B$ is given by a 6-dimensional integral over
  auxiliary momenta [4]. Numerical calculations give
  \beq
  B=4.98\pm 0.01
  \eeq
  in agreement with the analytic result recently obtained in [7,8]
  (their $g_{3P}=(2\pi)^4B$).

  Putting explicit expressions for $P$'s and $\gamma$'s into (20)
  and performing the integration over $s_1$ one finds 
  \beq
  a_2=-c_2\frac{R^2}{Q}\frac{s^{2\Delta}}{(\beta\ln s)^3}
  \frac{\ln^2\kappa}{\kappa}
  \eeq
  where $c_2$ is a known number.
  From this purely asymptotic expression one could  come to two
  conclusions. First, the double
  scattering contribution 
  behaves in $Q$ exactly as the single scattering term, that is, as $1/Q$.
  As a result, the EMC effect related to it should be independent
  of $Q$. Of course, this is an immediate consequence of the factorization
  property of the asymptotic triple pomeron interaction.
  Second, the singularity at small $\kappa$  leads to a stronger dependence
  on $A$. Since roughly speaking  $\kappa\sim 1/R_A$ where $R_A$ is
  the nuclear
  radius, from (27) one could conclude that the double scattering is
  enhanced by a factor $(R_A/R)\ln^2(R_A/R)$ as compared to the standard
  eikonal result.

  However all these conclusions are in fact wrong, since the upper pomeron
  enters the triple pomeron interaction not at asymptotic energies
  $\alpha_s\ln s>>1$ but at lower ones $\alpha_s\ln s\sim1$, where the
  asymptotic expression for its Green function  used in (20) is not valid.
  So we have to recur to the exact expression for it. Due to azimuthal symmetry
  of the projectile colour density we can retain only terms with zero
  orbital momentum in it:
\beq
G_{0}(s,r,r')=(1/8)rr'\int_{-\infty}^{\infty}\frac{d\nu s^{\omega(\nu)}}
{(\nu^{2}+1/4)^{2}}(r/r')^{-2i\nu}
\eeq
where
\beq
\omega(\nu)=2(\alpha_sN/2\pi)(\psi(1)-{\rm Re}\psi(1/2+i\nu))
\eeq
If we take this expression, put it into $\chi_1$, Eq. (10) and integrate
first over $s_1$, as indicated in (10), and afterwards over $r$, we obtain
\beq
\int ds_1 s_1^{-1-2\Delta}\chi(s_1,0,q)=-\frac{4\pi}{q}
\int  d^2r' r'\rho_1(r') I(q,r')
\eeq
where $I(q,r')$ is the remaining integral over $\nu$:
\beq
I(q,r')=\int d\nu\frac{(qr'/2)^{2i\nu}}
{2\Delta-\omega(\nu)}\frac{\Gamma(1/2-i\nu)}{\Gamma(1/2+i\nu}
\eeq

The asymptotic expression discussed above follows if one takes in (31)
all terms except the denominator $2\Delta-\omega(\nu)$ out of the
integral at $\nu=0$ and
in the denominator presents $\omega(\nu)=\Delta-\beta\nu^2$. Evidently
this procedure is wrong.
In fact, the integral (31) can be calculated as a sum of residues of the
integrand at
points $\nu=\pm ix_{k}$, $0<x_{1}<x_{2}<...$, at which
\beq
2\Delta-\omega(\nu)=0
\eeq
Residues in the upper semiplane are to be taken if $qr'/2>1$ and those
in the lower semiplane if $qr'/2<1$. Thus we obtain
\beq
I(q,r')=\frac{2\pi^{2}}{\alpha_sN}\sum_{k}c_k^{(\pm)}
(qr'/2)^{\pm 2x_{k}}
\eeq
where 
\beq
c^{(\pm)}_{k}=
\frac{\Gamma(1/2\mp x_{k})/\Gamma(1/2\pm x_{k})}
{\psi'(1/2-x_{k})-\psi'(1/2+x_{k})}
\eeq
and the signs should be chosen to always have $(qr'/2)^{\pm 2x_{k}}<1$.

The first three roots of Eq. (32) are
\beq 
x_{1}=0.3169,\ \ x_{2}=1.3718,\ \ x_{3}=2.3867
\eeq
with the corresponding coefficients $c_{k}^{(\pm)}$
\[
c_{1}^{(+)}=0.1522,\ \  
c_{2}^{(+)}=-0.1407,\ \  
c_{3}^{(+)}=0.03433,\ \ 
\]
\beq
c_{1}^{(-)}=0.007866,\ \  
c_{2}^{(-)}=-0.001802,\ \  
c_{3}^{(-)}=0.004494,\ \ 
\eeq

Returning to Eq. (5) for $a_2$ as an integral of the discontinuity
$D_1$ and
putting expressions for $\chi_2$ and the integrated $\chi_2$,
Eqs. (15) and (30), into the latter we obtain
\[
a_2=-128\pi^2\alpha_s^4N(N^2-1)\frac{s^{2\Delta}}{(\beta\ln s)^3}
F^2_2(\kappa)\sum_k\int d^2r r\rho_1(r)\]\beq
[c^{(+)}_k(\frac{r}{2})^{2x_k}B^{(+)}_k(r,\kappa)+
c^{(-)}_k(\frac{r}{2})^{-2x_k}B^{(-)}_k(r,\kappa)]
\eeq
where
\beq
B_k^{(\pm)}(r,\kappa)=\int\frac{d^2q}{(2\pi)^2q}q^{\pm 2x_k}
J^2(\kappa,q-\kappa/2)
\theta (\pm(2/r-q))
\eeq

As we observe, in the general case the factorization property is lost:
the integrals $B^{(\pm)}$ depend nontrivially both on the projectile and
target variables. However one can see that this property is restored in
the  limit of high $Q^2$, relevant for the hadronic structure functions.
In fact, in this limit the characteristic values of $r$ are small:
$r\sim1/Q$. Let us study how the integrals $B^{(\pm)}$ behave at small
$r$. In $B^{(\pm)}$ evidently large values of $q$ are essential.
The integrals $J$ behave as $\ln q/q$ at $q\rightarrow\infty$.
This leads to the following behaviour. 
\[B_k^{(+)}(r,\kappa)\sim r^{1-2x_k}\ {\rm if}\ 2x_k>1\ {\rm and}\
\sim const\ {\rm if}\  2x_k<1\]
\[B_k^{(-)}(r,\kappa)\sim r^{1+2x_k}\]
Combining this with other factors depending on $r$ we see that all terms
multiplying $\rho_1$ in the integrand behave as $r^2$ at small $r$, except
the first term with $k=1$, which, due to $2x_1<1$,  behaves as $r^{1+2x_1}$.
Evidently this term gives the dominant contribution in the limit
$Q^2\rightarrow\infty$, when (37) simplifies to
\beq
a_2=-2^{7-2x_1}\pi^2\alpha_s^4N(N^2-1)c^{(+)}_1
\frac{s^{2\Delta}}{(\beta\ln s)^3}
F^2_2(\kappa)\frac{b_1B_1}{Q^{1+2x_1}\kappa^{1-2x_1}}
\eeq
where the numbers $b_1$ and $B_1$ are very similar to our old $b$ and $B$
with additional powers of the variable in the integrand:
\beq
b_1=Q^{1+2x_1}\int d^2r r^{1+2x_1}\rho_1(r)
\eeq
(it does not depend on $Q$) and
\beq
B_1=\kappa^{1-2x_1}\int\frac{d^2q}{(2\pi)^2q}q^{2x_1}J^2(\kappa,q-\kappa/2)
\eeq
(it does not depend on $\kappa$).
Numerical calculations give
\beq
b_1^{(T)}=0.3145\,e^2Z^2,\ b^{(L)}=0.04377\,e^2Z^2,\ B_1=17.93
\eeq

Thus in the high-$Q$ limit the expression for $a_2$ fully factorizes
in the projectile and target. Its dependence on $Q$ and $\kappa$ turns
out to be intermediate between the eikonal and asymptotic triple pomeron
predictions. It vanishes at large $Q$ as $1/Q^{1+2x_1}$, faster than
the single pomeron
exchange and asymptotic triple pomeron ($\sim1/Q$) but not so fast as the
eikonal prediction $1/Q^2$. It is also singular at $\kappa\rightarrow 0$,
but the singularity is weaker than predicted by the asymptotic triple
pomeron.

The  double scattering amplitude following from (39) is
\beq
{\cal A}_2=-iC_A^22^{11-2x_1}\pi^2
\alpha_s^4\frac{N^2-1}{N}c^{(+)}_1 \tilde{w}_2
b_1B_1\frac{s^{2\Delta}}{(\beta\ln s)^3}
\frac{R^2}{(QR_A)^{1+2x_1}}
\eeq
where with a logarithmic accuracy (see (18))
\beq
\tilde{w}_2=R_A^{1+2x_1}\int\frac{d^2\kappa}{(2\pi)^2}
\kappa^{-1+2x_1}F_A(\kappa)\ln^2\kappa
\eeq
and we have introduced the nuclear radius $R_A$ to make $\tilde{w}_2$
dimensionless.
Passing to the low-$x$ EMC effect we have to compare it with the
dominant single scattering term
  \beq
  {\cal A}_1=iC_A^1\gamma_1\gamma_2(0)P(s,0)
  \eeq
  where, calculated at $\kappa$ exactly equal to zero, $\gamma_2(0)$ is
  finite:
  \beq
  \gamma_2(0)=2\alpha_s\sqrt{N^2-1}(R/N)
  \eeq
  The EMC effect is characterised by the EMC ratio $R_{EMC}$ of the nuclear
  structure function to $A$ times the nucleon one.
  The double scattering contribution   to it is  given by
  $R_{EMC}=1-\lambda_A$ where $\lambda_A=-{\cal A}_2/{\cal A}_1$.
  From our formulas we
  find
   \beq
   \lambda_A=c(A-1)\tilde{w}_2\frac{R}{R_A}\frac{1}{(QR_A)^{2x_1}}
   \frac{s^{\Delta}}{(\beta\ln s)^{5/2}}
    \exp\left(\frac{\ln^2Q}{\beta\ln s}  \right)
   \eeq
   where the numerical coefficient is
\beq
   c=2^{9-2x_1}\pi^{3/2}\alpha_s^2c_1^{(+)}B_1(b_{1T}+b_{1L})/(b_T+b_L)
\eeq

>From (47) we expect the EMC effect to go to zero as $Q\rightarrow\infty$
as $Q^{-2x_1}$, that is, rather slowly, essentially slowlier than one
would find from the eikonal picture. Its dependence on $A$ is enhanced
 by  a factor $\sim (R_A/R)^{1-2x_1}
\ln^2(R_A/R)$ as comparted to the eikonal prediction. The enhancement is
not so strong as one might naively predict on the basis of the
asymptotic triple pomeron picture.

\section{Some numerical estimates}
 The obtained formulas in principal allow to calculate the double
scattering contribution to the EMC effect on nuclei at small $x$.
Since the EMC effect is small experimentally, one may hope that this
contribution practically exhausts it (also see Conclusions for some
justification). Our formulas contain two parameters:
the strong coupling constant $\alpha_s$ and the nucleon radius $R$. The
value of $\alpha_s$ can be extracted from the observed intercept $\Delta$.
As to $R$, it has to be of the order of the proton electromagnetic radius.
Of course, one has to take into account inevitable uncertainties
associated with the logarithmic character of the hard pomeron model and
absence  of scale in $\log s$ and $\log \kappa$. However in trying to apply
our formulas to the experimental situation one meets with another serious
obstacle.

Comparison of the experimental proton structure function with the predictions
based on the hard pomeron model indicates that the model can only be valid
at very small $x\leq x_0=0.01$. A power growth characteristic for the
hard pomeron can be supported experimentally at
$x\sim x_1=10^{-4}$ - $10^{-5}$.
The triple pomeron interaction picture requires that both $s_1$ for the upper
pomeron and $s/s_1$ for the two lower ones be correspondingly large.
It follows that this picture can only be applicable at extraordinary small
$x\leq x_0x_1=10^{-6}$.
Besides,the large $Q^2$ limit used to obtain a factorizable form for the
contribution requires that $(QR_A)^{1-2x_1}>>1$. Present experimental data
on the EMC effect do not satisfy these requirements. They are restricted
to $x>0.005$ and values of $Q^2$ below 2 $(GeV/c)^2$ for $x<0.01$.
Thus,
 comparison of our predictions to the existing data cannot be
justified.

Just for curiosity, if, notwithstanding this objection, one takes
realistic values 
$\Delta=0.2$ and $R=0.44\ fm$ [9], then one gets for Ca ($A=40$) at $x=0.0085,\
Q^2=1.4\, (GeV/c)^2$
$\lambda_{40}=0.14$ in an  agreement with the experimental value
$\lambda_{40}^{ex}=0.154\pm 0.014$. However with the same parameters one gets
for Xe ($A=131$) at $x=0.0065,\ Q^2=1.34\,(GeV/c)^2$ and C
($A=12$) at $x=0.0055,\ Q^2=1.1\,(GeV/c)^2$) values
$\lambda_{131}=0.38$ and $\lambda_{12}=0.043$ compared to the experimental
values $\lambda_{131}^{ex}=0.16\pm 0.11$ and
$\lambda_{12}^{ex}=0.096\pm 0.013$.

Forgetting about the experimental situation,
our predictions for the range of $x$ and $Q^2$ where one  can expect the
obtained formulas  to be applicable are illustrated in the Table.
One observes from it that the dependence on $x$ is quite weak.
In fact one obtains practically the same values of $\lambda$
in the whole range $10^{-6}<x<10^{-2}$ due to compensation of the growth
with $1/x$ of both the numerator and  denominator in  (47). 
The predicted dependence on $Q^2$ is clearly visible, although also
not very strong as follows from (47).

\section{Conclusions}
 Two main consequences follow for the double scattering contribution
to the structure functions of the nuclei from the hard pomeron model.
First, at large $Q^2$ it behaves as $Q^{-1-2x_1}$ where
$x_1=0.3169$ is a number which does not depend on the coupling and so
is exactly
known. As a result the corresponding low-$x$ EMC effect should go to zero
as $Q^{-2x_1}$. This property does not seem to be changed after
the inclusion of higher order
rescatterings. Indeed in the dipole approach one finds that they are
described by the pomeronic fan diagrams [4]. Then the initial pomeron
coupled to the projectile will always be in the same regime as in our double
scattering case with the only difference that the rest part
will
behave as $S^{n\Delta}$ for $n$ rescatterings.
The power $x_1^{(n)}$ governing the high $Q^2$ behaviour will be
the smallest
positive root of the equation (32) with $2\Delta\rightarrow n\Delta$.
This power grows with $n$: $x_1^{(3)}=0.3793,\ x_1^{(4)}=0.4097,\ ...
,\,x_1^{(\infty)}=0.5$.  Contributions from higher order rescatterings
will then go down 
at high $Q^2$ faster than the double  scattering contribution,
although by a rather small power of $Q^2$. 
 Thus  the behaviour $Q^{-2x_1}=Q^{-0.6338}$ seems
to be a clear prediction for the small-$x$ EMC effect which may serve to
test the hard pomeron model.

Second, due to singularities at small momentum transfers, the double
scattering turns out to be more strongly dependent on $A$ as compared to the
standard eikonal predictions. This second prediction does not, however, seem
so trustworthy, since it relies on the specific property of the lowest order
hard pomeron model of having no intrinsic scale.

\section{References}
1. A.Mueller, Nucl. Phys.,{\bf B415} (1994) 373.\\
2. J.Bartels and M.Wuesthoff, Z.Phys., {\bf C66} (1995) 157.\\
3. R.Peschanski, Phys. Lett. {\bf B409} (1997) 491.\\
4. M.A.Braun and G.P.Vacca, Bologna univ. preprint, hep-ph/9711486.\\
5. N.N.Nikolaev and B.G.Zakharov, Z. Phys. {\bf C49} (1991) 607.\\
6. M.A.Braun, Z.Phys. {\bf C71} (1996) 123.\\
7. G.P.Korchemsky, preprint LPTHE-Orsay-97-62; hep-ph/9711277.\\
8. A.Bialas, H.Navelet and R.Peschanski, Saclay preprints
hep-ph/9711236  and hep-ph/9711442.\\
9. N.Armesto and M.A.Braun, Z.Phys. {\bf C76} (1997) 81. \\                       \\

\newpage
\begin{center}
{\Large{\bf Table}}\\
\vspace{0.5 cm}
Predictions for the EMC ratios at large$1/x$ and $Q^2$.

\vspace{1.5 cm}

\begin {tabular}{|r|c|c|c|}\hline
   $A$ & $x$ & $Q^{2}$ (GeV$^{2}$)&   $1-R_{EMC}$\\\hline
       40 & $10^{-5}$&  10.  &  0.0685  \\
          &          &  100. &  0.0330  \\
          & $10^{-6}$&  10.  &  0.0775  \\
          &          &  100. &  0.0374  \\\hline
       64 & $10^{-5}$&  10.  &  0.104   \\
          &          &  100. &  0.0502  \\
          & $10^{-6}$&  10.  &  0.118   \\
          &          &  100. &  0.0568  \\\hline
      131 & $10^{-5}$&  10.  &  0.190   \\
          &          &  100. &  0.0914  \\
          & $10^{-6}$&  10.  &  0.215   \\
          &          &  100. &  0.103   \\\hline
\end{tabular}
\end{center}
\end{document}